\documentclass[aps,pra,twocolumn,showpacs,nofootinbib,longbibliography,notitlepage]{revtex4-2}
\usepackage{etex}
\usepackage{amsmath,amssymb,amsthm}
\usepackage[colorlinks=true,citecolor=blue,urlcolor=blue]{hyperref}
\usepackage[pdftex]{graphicx}
\usepackage{times,txfonts}
\usepackage{braket}
\usepackage{color}
\usepackage{natbib}
\usepackage{amsmath,blkarray}
\usepackage{mathtools}
\usepackage{latexsym}
\usepackage{tabularx, booktabs}
\usepackage{graphics,epstopdf}
\usepackage{graphicx}
\usepackage{float}
\usepackage{graphicx}
\usepackage{amsfonts}
\usepackage{subcaption}
\usepackage{color,soul}

\newcommand{\be}{\begin{equation}}
	\newcommand{\ee}{\end{equation}}
\newcommand{\ba}{\begin{eqnarray}}
	\newcommand{\ea}{\end{eqnarray}}

\begin{document}
	\title {Discriminating mirror symmetric states with restricted contextual advantage}
	
	\author{Sumit Mukherjee}
	\email{mukherjeesumit93@gmail.com}
	\author{Shivam Naonit}
	\email{shivamn.ug17.ph@nitp.ac.in}
	
	\author{ A. K. Pan }
	\email{akp@nitp.ac.in}
	\affiliation{National Institute of Technology Patna, Ashok Rajpath, Patna, Bihar 800005, India}
	
	\begin{abstract}
		The generalized notion of noncontextuality provides an avenue to explore the fundamental departure of quantum theory from a classical explanation. Recently, extracting a different form of quantum advantage in various information processing tasks has received an upsurge of interest. In a recent work [D. Schmid and R. W. Spekkens, \href{https://journals.aps.org/prx/abstract/10.1103/PhysRevX.8.011015}{Phys. Rev. X 8, 011015  (2018)}
		] it has been demonstrated that discrimination of two nonorthogonal pure quantum states entails contextual advantage when the states are supplied with equal prior probabilities. We generalized the work to arbitrary prior probabilities as well as to three arbitrary mirror-symmetric states. We show that the contextual advantage can be obtained for any value of prior probability when only two quantum states are present in the task. But surprisingly, in the case of three mirror-symmetric states, the contextual advantage is available only for a restrictive range of prior probabilities with which the states are prepared.
	\end{abstract}
	\maketitle
	\section{Introduction}
	
	The existence of nonclassicality in the fundamental description of nature manifests in several ways. Bell's theorem \cite{bell,chsh} and Kochen-Specker(KS) theorem \cite{KSpeker68,bell66} are the two most important discoveries which introduce empirically testable notions of nonclassicality, namely nonlocal causality and contextuality, respectively. Both of them provide evidence that some particular class of realistic descriptions (called hidden variable models \cite{harrigen}) of quantum theory are forbidden. But, unlike Bell's theorem \cite{chsh} involving the impossibility proof of local realistic description of quantum theory, the KS-theorem \cite{KSpeker68} is not formulated in a theory-independent manner. The requirement of measurement context while specifying deterministic values to observables was later generalized in \cite{spekkens05} for any arbitrary operational theory. We shall use the terminology 'classical' and 'noncontextual' interchangeably, and unless stated otherwise, by the terminology noncontextuality,' we refer to the generalized notion of it as provided in \cite{spekkens05}.
	
	The problem of discriminating between two quantum states has a fundamental importance in quantum foundations and information theory. While two nonorthogonal quantum states can not be perfectly discriminated \cite{helstrom}, there has been a multitude of attempts to find out the optimal strategies that can discriminate two states with at least a certain nonzero success probability. Some of the most successful strategies include unambiguous discrimination \cite{undis1,undis2,undis3} and minimum error discrimination \cite{errdis,discrimination02} of quantum states. The former is used when an accurate guess of the state is important, while the latter is employed to minimize the chance that an incorrect guess will occur. For our purpose, here we will keep our discussion on minimum error state discrimination (MESD) \cite{discrimination02}. In particular, we analyze MESD for mirror-symmetric states in quantum theory and examine whether a noncontextual ontological model (specified later) performs better in such a task.
	
	Over the last decade, a surge of works has been carried out with different purposes where the idea of operationalizing the MESD was explored \cite{maroni13,cavalcanti14,cyril14,ContxDis18}. In particular, such an investigation in terms of noncontextuality was first put forward in \cite{ContxDis18}, where the authors considered the MESD of two arbitrary nonorthogonal states supplied with equal prior probabilities. Using the noncontextual model, they formulated an interesting and empirically testable no-go theorem for MESD by providing an upper bound to the success probability and demonstrated that the quantum theory always outperforms the noncontextual model in this task. But it remains an open question whether the quantum advantage is available when the states are supplied with arbitrary prior probabilities and when more states are involved.
	
	In this work, by extending the formulation in \cite{ContxDis18} we investigate the MSED task for three mirror-symmetric states, supplied with arbitrary prior probabilities. We first show that in the case of MESD for two nonorthogonal states for any given prior probability, quantum theory always outperforms any noncontextual model. We then demonstrate that for MESD of three mirror-symmetric states, the quantum advantage can be achieved only restrictively. We explicitly show that for a specific set of three mirror-symmetric states, the availability of quantum advantage depends on the prior probabilities with which the states are supplied.
	
	This paper is organized as follows. In Sec. II, we discuss quantum strategy for MESD of two nonorthogonal pure states, and by briefly recapitulating the notion of noncontextuality, we discussed the noncontextual strategy for MESD. Section III discusses optimal strategies for both quantum and noncontextual models and compares their performances. In conclusion, we have discussed the implications of our result and future directions in Sec. IV.
	
	\section{MESD and Preparation non-contextuality }
	
	\textit{Quantum strategy for MESD:} It is a well established fact in quantum theory that Discriminating two orthogonal pure quantum states is a trivial task, it is not in general possible to discriminate between two non-orthogonal pure quantum states. For an example let us take states $\ket{\psi_{1}}$ and its orthogonal state $\ket{\bar{\psi_{1}}}$; this two quantum states can be discriminated perfectly (with probability one) by employing a measurement in the basis $\{\ket{\psi_{1}}\bra{\psi_{1}},\ket{\bar{\psi_{1}}}\bra{\bar{\psi_{1}}}\}$. But if one is interested to discriminate between two nonorthogonal states say $\ket{\psi_{1}}$ and $\ket{\psi_{2}}$ (where $\braket{\psi_{1}|\psi_{2}}>0$) by employing a measurement in the basis $\{\ket{\psi_{1}}\bra{\psi_{1}},\ket{\bar{\psi_{1}}}\bra{\bar{\psi_{1}}}\}$ or $\{\ket{\psi_{2}}\bra{\psi_{2}},\ket{\bar{\psi_{2}}}\bra{\bar{\psi_{2}}}\}$ then there is a certain probability that $\ket{\psi_{2}}$ would pass the test for $\ket{\psi_{1}}\bra{\psi_{1}}$ and the vice-versa. This will constitute an error in terms of the degree of overlap between the quantum states. We will call the quantity $c^{\psi_{1},\psi_{2}} =|\braket{\psi_{1}|\psi_{2}}|^{2}$ as confusability as it is the probability that one makes a mistake by guessing the wrong states. Among a few successful strategies of discriminating two pure quantum states, we are interested in a particular one known as minimum error state discrimination. In such a strategy, one is interested in minimizing the error probability while discriminating the states. The basic scenario is as follows. A set of possible states $\{\psi_{\tiny i}\}$ is supplied with a prior probability $p_{\tiny i}$ and the job of the experimenter is to determine which state is supplied in a particular run. The measurement strategy for discrimination of two non-orthogonal pure states $\ket{\psi_{1}}$ and $\ket{\psi_{2}}$ involves finding a two outcome measurement characterized by POVMs $\{E_{\psi_{1}},E_{\psi_{2}}\}$ so that the success probability becomes,

	\begin{equation}
		S_{2}= p_{1}Tr[\ket{\psi_{2}}\bra{\psi_{2}}E_{\psi_{2}}]+p_{2}Tr[\ket{\psi_{1}}\bra{\psi_{1}}E_{\psi_{1}}]
	\end{equation}
	where p$_{1}$ and p$_{2}$ are prior probabilities with which the states are supplied. In minimum error state discrimination the task of the experimenter is to minimize the error probability E$_Q$=(1-$S_{2})$.
	Helstrom \cite{helstrom} found the optimum success probability by finding out the optimum POVMs which minimizes the error probability and is given by,

	\begin{equation}
		S_{2} = \frac{1}{2}\left(1 + \sqrt{1-4p_{\tiny 1}p_{\tiny 2}c^{\psi_{1},\psi_{2}}}\right)\label{eq:gensucc}
	\end{equation}
	where $c^{\psi_{1},\psi_{2}}=\braket{\psi_{1}|\psi_{2}}|^{2}$ is the confusibility defined earlier. A nonzero value of confusibility means that there is no way of determining with perfection which of the two states a given physical system has been prepared.

	If the states are equally probable, that is, $p_{\tiny 1}$=$p_{\tiny 2}$=$\frac{1}{2}$, then the probability of guessing the state correctly using the Helstrom measurement is,
	\begin{equation}
		\begin{split}
			S_{2}=\frac{1}{2}\left(1 + \sqrt{1-c^{\psi_{1},\psi_{2}}}\right)
		\end{split}\label{eq:trade2}
	\end{equation}

	Here we recall the framework of the ontological model \cite{harrigen,spekkens05} of an operational theory to introduce the notion of noncontextuality from a modern perspective. But before going into the discussion of an ontological model, let us first describe what an operational theory is. The basic building blocks of an operational theory essentially consist of a set of preparation procedures, a set of transformations, and measurement procedures. For a given preparation procedure $P$ and a measurement procedure $M$, an operational theory provides statistics given by the probability $p(k|P, M)$ of obtaining a particular outcome $k$. The preparation procedures in quantum theory provide the density matrix $\rho$. The measurement procedures are represented as positive semidefinite operators, i.e., the positive operator valued measures or POVMs (E$_{k}$). Forgiven quantum state and measurement operator Born rule provides the statistical predictions for the outcomes of an experiment as $p(k|P, M)=Tr[\rho E_{k}]$.\\

	\textit{Preparation noncontextuality:} Two preparations P$_{0}$ and P$_{1}$ are said to be operationally equivalent if any experimental procedures can not distinguish them.  The same density matrices describe equivalent preparations in quantum theory. Similarly, two measurements are operationally equivalent in quantum theory if identical operators represent them. But to avoid unnecessary discussions, we will restrict ourselves to the preparation procedures only.

	An ontological model \cite{ballentine,harrigen} of an operational theory is formulated to provide an objective description of the experimental statistics. In an ontological model  a ontic state corresponding to a preparation procedure P is described by a probability distribution $\mu(\lambda|P)$ of an unknown variable $\lambda \in \Lambda$ which follows the following completeness relation, $\int_{\Lambda}\mu(\lambda|P) d\lambda =1$.
	
	So an ontological model that is compatible with quantum theory must reproduce the Born rule for the outcome probabilities such that,
	
	\begin{equation}
		\int_{\Lambda}\mu(\lambda|\rho)  \xi_{M}(\phi_{k}|\lambda) d\lambda = Tr(\rho E_{k}), \label{eq:born}
	\end{equation}

	where $E_{k}= \ket{\phi_{k}}\bra{\phi_{k}} $ is the projector corresponding to the k$^{th}$ outcome of measurement M and $\xi_{M}(\phi_{k}|\lambda)$ is the corresponding response function in ontological model, such that, $\forall\lambda$ $\sum_{k}\xi_{M}(\phi_{k}|\lambda)=1$.
	
	Although the ontological model for the quantum theory that we are looking for is committed to reproducing the Born rule, it is not guaranteed that a certain restriction on the theory's operational level would reflect the respective probability distributions over the ontic state variable $\lambda$. This inspires one to define the notion of preparation noncontextuality defined in Ref. \cite{spekkens05} and is represented as follows. If two operationally equivalent preparations say P$_{1}$ and P$_{2}$ in a theory is represented by a equal ontic state probability distribution in corresponding ontological model, viz., $P_{0}\approx P_{1}	\Rightarrow \mu(\lambda|P_{0})=\mu(\lambda|P_{1})$, then the model is said to be preparation noncontextual. In quantum theory if preparation procedure P$_{1}$ and P$_{2}$ prepare identical states represented by same density operators then the preparation noncontextuality assumption implies that the ontic state probability distributions corresponding to those two density operators are same such that, $p(k|P_{0}, M)=p(k|P_{1},M)	\Rightarrow \mu_{P_{0}}(\lambda|\rho)=\mu_{P_{1}}(\lambda|\rho)$.

	\textit{Noncontextual strategy for MESD:} It is a well-known fact that preparation contextuality provides advantages in many information processing tasks\cite{ghorai18,spekk09,pan19,saha19}. Recently, Schmid and Spekkens \cite{ContxDis18} have investigated whether the presence of contextuality in the ontological description of quantum theory provides an advantage in the MESD task. They have started with the generic problem of discriminating two nonorthogonal pure quantum states represented on the great circle of a Bloch sphere. A general operational theoretic version of the task facilitates an ontic description of states embedded with the assumption of preparation noncontextuality. As we know, that the MESD is about guessing the right state in each run of an experiment. The analog of the task in an ontological model is about guessing which distribution an ontic state variable $\lambda$ comes from. Thus, an ontic state variable $\lambda$ is sampled with a certain probability p($\lambda$) from one of the two overlapping probability distributions, $\mu_{\tiny \psi}(\lambda)$ and $\mu_{\tiny \phi}(\lambda)$. The guess should be such that it has the highest probability of being correct. So for a certain $\lambda$ the maximum probability of distinguishing the preparations is max$\{\mu(\lambda|\psi),\mu(\lambda|\phi)\}$. Also from Bayesian inversion if $\mu(\lambda|\psi)>\mu(\lambda|\phi)$ then $\mu(\psi|\lambda)>\mu(\phi|\lambda)$. So the probability of guessing the correct distribution is simply $\{\mu(\psi|\lambda),\mu(\phi|\lambda)\}$. On average then, the success probability S$_{ont}$ can be written as,
	\begin{equation}
		\begin{split}
			S_{2}&=\int_{\Lambda}p(\lambda) max\{\mu(\psi_{1}|\lambda),\mu(\psi_{2}|\lambda)\}d\lambda \\
			&= \int_{\Lambda} p(\lambda)\Big(1- min\{\mu(\psi_{1}|\lambda),\mu(\psi_{2}|\lambda)\}\Big)\\
			&= 1-\int_{\Lambda}\Big( min\{\mu(\lambda|\psi_{1})p(\psi_{1}),\mu(\lambda|\psi_{2})p(\psi_{2})\}\Big)d\lambda \label{eq:Sont1}
		\end{split}
	\end{equation}
	Any ontological model that respects the asssumption of preparation noncontextuality ends up giving the bound \cite{ContxDis18} of maximum success probability $S_{2}$ given by ,
	\begin{equation}
		\begin{split}
			S_{2}\leqslant 1-\frac{1}{2}c^{\psi_{1},\psi_{2}}\label{eq:trade1}
		\end{split}
	\end{equation} 
	Therefore, ref.\cite{ContxDis18} has established a no-go theorem in terms of the trade-off between the success probability $S_{2}$ and confusiblility for a noncontextual model. It is thereby evident from Eq. \eqref{eq:trade2} and Eq. \eqref{eq:trade1} that quantum strategy for dicriminating two arbitrary pure quantum states allows higher success rates for a given confusiblility than its preparation noncontextual counterpart. 
	
	We note here that the authors in  \cite{ContxDis18} have considered a scenario where the states are supplied with equal prior probability. So a natural question may arise that is whether it is true for any arbitrary prior probabilities. The question is answered through the following proposition followed by formal proof.

	From Eq.  \eqref{eq:Sont1} the success probability S$_{2}$  can be re-written in terms of the following inequality ,
	\begin{equation}
		S_{2}\leq 1- min\{p(\psi_{1}),p(\psi_{2})\} \int_{\Lambda} min\{\mu(\lambda|\psi_{1}),\mu(\lambda|\psi_{2})\}d\lambda,
	\end{equation}
	where we have used the following inequality,
	\begin{equation}
		\begin{split}
			\int_{\Lambda}\Big( &min\{\mu(\lambda|\psi_{1})p(\psi_{1}),\mu(\lambda|\psi_{2})p(\psi_{2})\}\Big)d\lambda\\
			&\geq min\{p(\psi_{1}),p(\psi_{2})\} \int_{\Lambda}\Big( min\{\mu(\lambda|\psi_{1}),\mu(\lambda|\psi_{2})\}\Big)d\lambda
		\end{split}
	\end{equation}
	If we consider the prior probabilities to be p($\psi$)= p and p($\phi$)= 1-p then  $min\{p(\psi_{1}),p(\psi_{2})\}=p(\psi_{1})=p$ for  $p\leqslant 1/2$  and $min\{p(\psi_{1}),p(\psi_{2})\}=p(\psi_{2})=1-p$ for  $p> 1/2$. The success probability S$_{2}$ can be written as,\\
	$\forall \lambda,\psi_{i}$.	
	
	for $p\leqslant 1/2$
	\begin{equation}
		\begin{split}
			S_{2} \leqslant 1-p\int_{\Lambda} min\{\mu(\lambda|\psi_{1}),\mu(\lambda|\psi_{2})\}d\lambda \nonumber
		\end{split}
	\end{equation}

	for $p> 1/2$
	\begin{equation}
		\begin{split}
			S_{2} \leqslant 1-(1-p)\int_{\Lambda} min\{\mu(\lambda|\psi_{1}),\mu(\lambda|\psi_{2})\}d\lambda \label{eq:succ1}\\
		\end{split}
	\end{equation}

	We define the confusibility between quantum preparations $\ket{\psi_{1}}$ and $\ket{\psi_{2}}$ in an ontological model by the overlap between the corresponding ontic state probabilities $\mu(\lambda|\psi_{1})$ and $\mu(\lambda|\psi_{2})$ and is given by,
	\begin{eqnarray}
		c^{\psi_{1},\psi_{2}}&&=\int_{\Lambda}\mu(\lambda|\psi_{1})  \xi_{M_{\psi_{2}}}(\psi_{2}|\lambda) d\lambda  \nonumber\\
		&&= \int_{\Lambda_{\psi_{2}}}\mu(\lambda|\psi_{1})   d\lambda +\int_{\Lambda/\Lambda_{\psi_{2}}}\mu(\lambda|\psi_{1})  \xi_{M_{\psi_{2}}}(\psi_{2}|\lambda) d\lambda
		\label{eq:conf}
	\end{eqnarray} \\

	In the first integral of Eq. \eqref{eq:conf} we have used the fact that $\forall\lambda\in\Lambda_{\psi_{2}}$, $ \xi_{M_{\psi_{2}}}(\psi_{2}|\lambda)=1$. In general, the second term of Eq. \eqref{eq:conf} can have a nonzero value but we will prove shortly that this integral is zero if the ontological model is preparation noncontextual. \\
	
	\textbf{Lemma:} \textit{In a preparation noncontextual ontological model of quantum theory for any two nonorthogonal pure states $\ket{\psi_{1}}$ and $\ket{\psi_{2}}$ the quantity $\mu(\lambda|\psi_{1})  \xi_{M_{\psi_{2}}}(\psi_{2}|\lambda)$ has null support for all $\lambda\in\Lambda/\Lambda_{\psi_{2}}$ }.\\
	
	A maximally mixed state is prepared in infinitely many ways. For an example let us take the states $\ket{\psi_{1}}$, $\ket{\psi_{2}}$ and their orthogonal states $\ket{\bar{\psi_{1}}}$ and $\ket{\bar{\psi_{2}}}$ respectively which in convex decomposition gives maximally mixed state as $\big(\ket{\psi_{1}}\bra{\psi_{1}}+\ket{\bar{\psi_{1}}}\bra{\bar{\psi_{1}}}\big)/2 =\big(\ket{\psi_{2}}\bra{\psi_{2}}+\ket{\bar{\psi_{2}}}\bra{\bar{\psi_{2}}}\big)/2=\mathbb{I}/2$. It is shown in ref. \cite{spekkens05} that the convex combination of preparations in the operational level is represented by the convex sum of the corresponding ontic state probabilities at ontic state level. Taking this into account the preparation noncontextuality assumptions for the maximally mixed state lead to the following mathematical description,
	
	\begin{equation}
		\frac{(\mu(\lambda|\rho_{\psi_{1}})+\mu(\lambda|\rho_{\bar{\psi_{1}}}))}{2}=\frac{(\mu(\lambda|\rho_{\psi_{2}})+\mu(\lambda|\rho_{\bar{\psi_{2}}}))}{2} = \mu\Big(\lambda|\frac{\mathbb{I}}{2}\Big) \label{eq:cntx}
	\end{equation}
	
	Let us take the two distribution $\mu(\lambda|\rho_{\psi_{1}})+\mu(\lambda|\rho_{\bar{\psi_{1}}})$)/2 and $\mu(\lambda|\rho_{\psi_{2}})+\mu(\lambda|\rho_{\bar{\psi_{2}}})$)/2 corresponding to two different preparation procedure has supports $\Lambda_{1}=\Lambda_{\psi_{1}}\cup\Lambda_{\bar{\psi_{1}}}$ and $\Lambda_{2}=\Lambda_{\psi_{2}}\cup\Lambda_{\bar{\psi_{2}}}$ respectively. Let us also consider the space $\Omega$ as the space where the second integral of Eq. \eqref{eq:conf} is nonzero and s given by $\Omega=\{\lambda\in\Lambda/\Lambda_{\psi_{2}}|\hspace{0.05cm}  \mu(\lambda|\psi_{1})>0 , \xi_{M_{\psi_{2}}}(\psi_{2}|\lambda)>0\}$. So the intersection space $\Omega\cap\Lambda_{1}$ has a nonzero support as $\mu(\lambda|\psi_{1})>0$ even outside the space $\Lambda_{\psi_{2}}$. But the intersection space $\Omega\cap\Lambda_{2}$ is zero as in the space $\forall\lambda\in\Lambda/\Lambda_{\psi_{2}}$, $\mu(\lambda|\psi_{2})=0$  and $\forall\lambda\in\Lambda_{\bar{\psi_{2}}}$, $\xi_{M_{\psi_{2}}}(\psi_{2}|\lambda)=0 $. But for a preparation noncontextual model $\Lambda_{1}=\Lambda_{2}$. So the space $\Omega$ cannot be nonzero in a preparation noncontextual model thereby yielding the second integral of Eq. \eqref{eq:conf} to be zero, which proves the Lemma.\\
	
	An important implication of the above Lemma is that for a preparation noncontextual ontological model of quantum theory the response function $\xi(\psi_{2}|\lambda)$ is outcome deterministic, i.e., $\forall \lambda \in supp[\mu(\lambda|\psi_{2})],\hspace{0.2cm} \xi(\psi_{2}|\lambda)=1$ and $\xi(\psi_{1}|\lambda)=0$ otherwise. This has been proved in ref. \cite{spekkens05} differently. Furthermore, using the Lemma in Eq. \eqref{eq:conf} we can write the confusibility $c_{NC}^{\psi_{1},\psi_{2}}$ for a noncontextual model  as,
	
	\begin{equation}
		c^{\psi_{1},\psi_{2}}=\int_{\Lambda_{\psi_{2}}}\mu(\lambda|\psi_{1}) d\lambda  \label{eq:conf1}
	\end{equation} 
	
	But as $\ket{\psi_{1}}$ and $\ket{\bar{\psi_{1}}}$ are orthogonal, they have zero support in the ontic state space i.e., $\mu(\lambda|\psi_{1})\mu(\lambda|\bar{\psi_{1}})=0$ and similarly $\mu(\lambda|\psi_{2})\mu(\lambda|\bar{\psi_{2}})=0$. So using this fact we get from Eq.\eqref{eq:cntx} $\forall\lambda\in\Lambda_{\psi_{1}}\cap\Lambda_{\psi_{2}}$, $\mu(\lambda|\psi_{1})=\mu(\lambda|\psi_{2})==2\mu(\lambda|\mathbb{I}/2)$. As we are interested in the space $\Lambda_{\psi_{2}}$, $min\{\mu(\lambda|\psi_{1}),\mu(\lambda|\psi_{2})\}=\mu(\lambda|\psi_{1})=\mu(\lambda|\psi_{2})$ for all $\lambda\in\Lambda_{\psi_{2}}\cap\Lambda_{\psi_{1}}$ otherwise, $min\{\mu(\lambda|\psi_{1}),\mu(\lambda|\psi_{2})\}=0$. So from Eq. \eqref{eq:conf1} we get,
	\begin{equation}
		\begin{split}
			c^{\psi_{1},\psi_{2}}&=\int_{\Lambda_{\psi_{2}}}\mu(\lambda|\psi_{1}) d\lambda \nonumber \\
			&=\int_{\Lambda}min\{\mu(\lambda|\psi_{1}),\mu (\lambda|\psi_{2})\}d\lambda \label{eq:nonc}
		\end{split}
	\end{equation}

	\begin{figure}[H]%
		\centering
		\subfloat[\centering Side view with the extreme end representing equal prior probability ]{{\includegraphics[width=6cm]{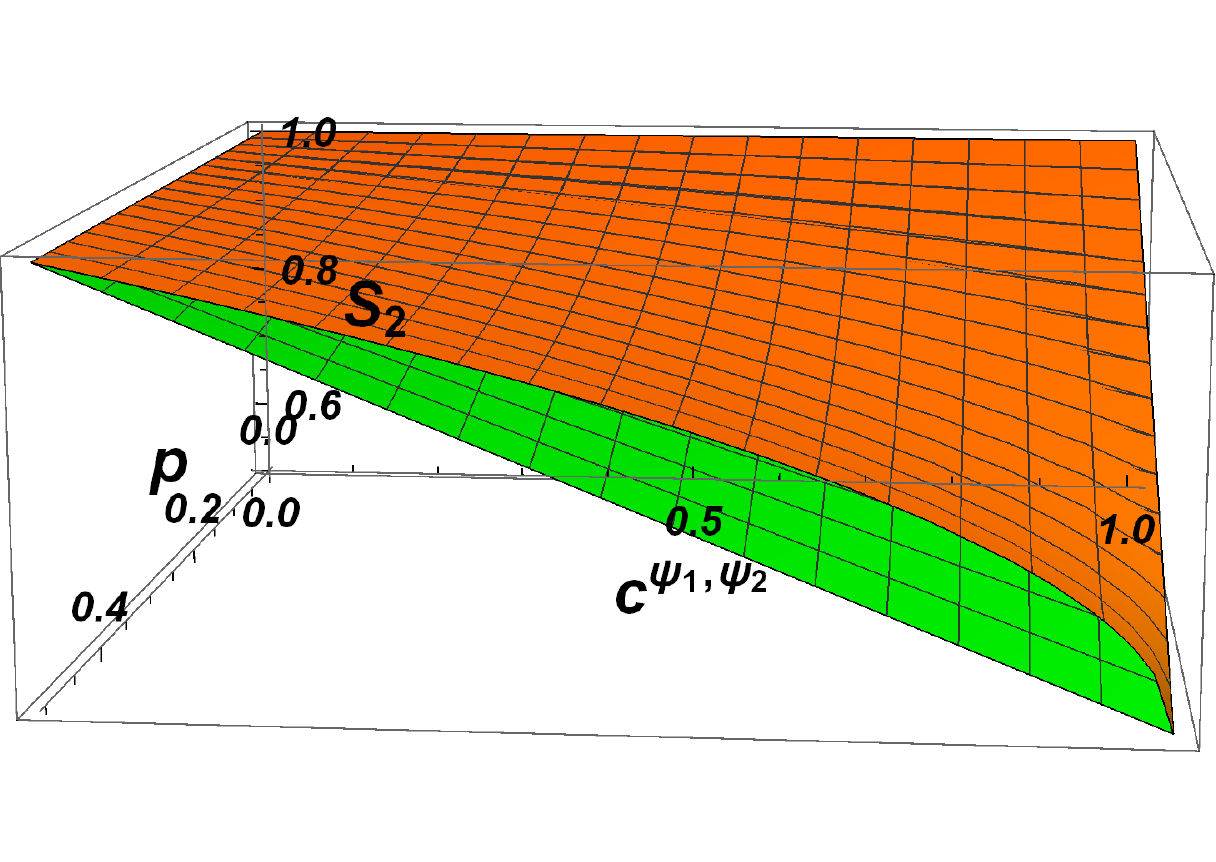} }}%
		\qquad
		\subfloat[\centering Front view]{{\includegraphics[width=5.5cm]{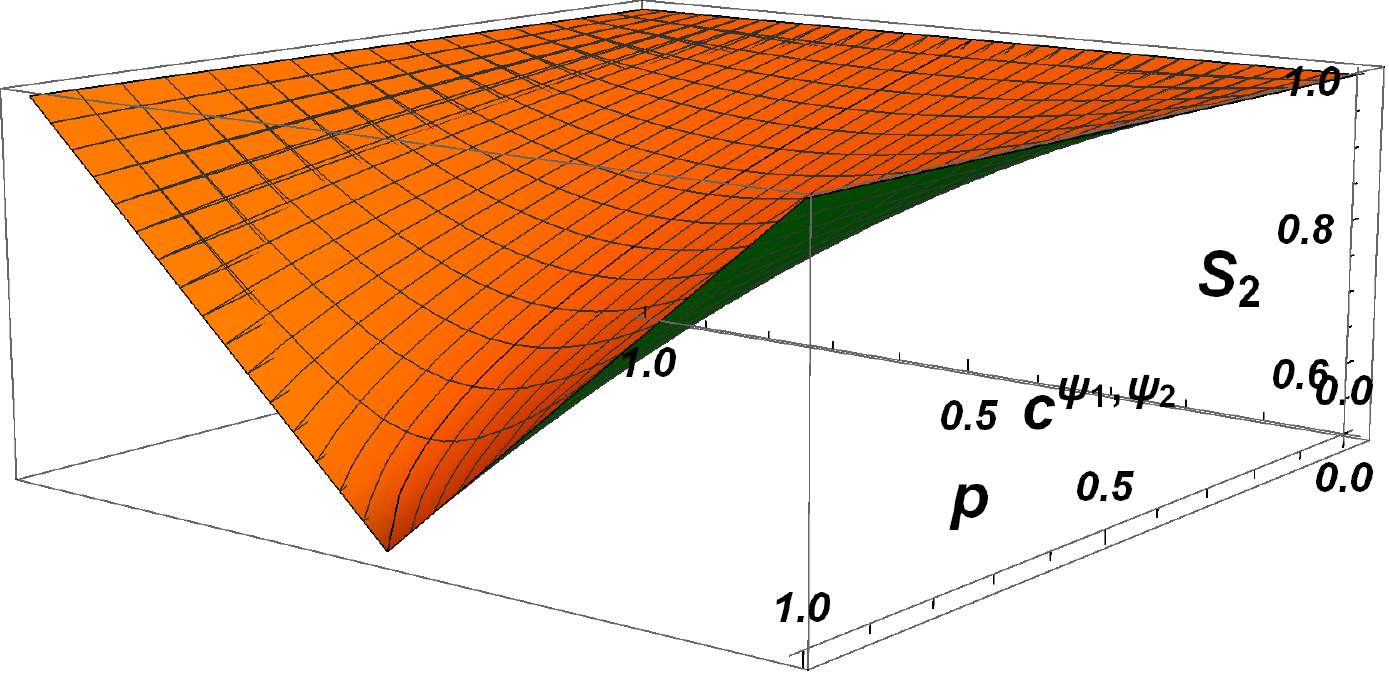} }}%
		\caption{Orange graph represents optimal trade of between the quantum success probability and confusibility for arbitrary prior probabilities while the Green graph is that for preparation noncontextual model.}%
		\label{fig:twoarb}%
	\end{figure}
	
	Impinging Eq. \eqref{eq:conf1} into Eq. \eqref{eq:succ1} we finally get,\\
	for $0\leq p\leqslant 1/2$
	\begin{equation}
		\begin{split}
			S_{2} \leqslant 1-pc^{\psi_{1},\psi_{2}}\nonumber
		\end{split}
	\end{equation}

	and for $1/2 < p \leq 1$
	\begin{equation}
		\begin{split}
			S_{2} \leqslant 1-(1-p)c^{\psi_{1},\psi_{2}} \label{eq:succ2}\\
		\end{split}.
	\end{equation}

	From Equation \eqref{eq:succ2} it is seen that the success probability $S_{2}$ is a function of $p$ and the confusability $c^{\psi_{1},\psi_{2}}$ and is plotted in Fig. \ref{fig:twoarb}. It is evident that even if the states are supplied with arbitrary prior probabilities, quantum theory always provides an advantage in a minimum error state discrimination task for two states. This leads us to write the following proposition.
	
	\textbf{Proposition:} \textit{For MESD of two nonorthogonal pure quantum states contextual advantage is available regardless of the prior probabilities with which the states are being supplied.}
	
	It is to be noted that the above proposition is limited in the sense that it is applied to the case when only two states are involved in the discrimination process. A more general case must investigate the scenario where more states are involved. Now we will consider in specific the case of MESD for three states. 
	
	\section{Optimal strategy for discriminating mirror-symmetric states}

Here we discuss the optimum strategy of discriminating three qubit states possessing a mirror symmetry. By the term mirror symmetry we mean that a transformation $\{\mathcal{T}\in O(\mathcal{H})\ \hspace{0.2cm}  |\hspace{0.2cm}  \mathcal{T}\ket{0}\rightarrow\ket{0},\mathcal{T}\ket{1}\rightarrow-\ket{1}\}$ leaves the set of states invariant. In minimum error discrimination the goal is to minimize the error probability. For example let us take a set of states $\{\ket{\psi_{i}}\}$ with prior probability $\{p_{i}\}$ are being sent to the detector where a for successful detection of state $\ket{\psi_{i}}$ detector D$_{i}$ should click. In quantum theory particular detector D$_{i}$ is represented by a Projective valued measure $\pi_{i}$. So the probability that a given state $\ket{\psi_{i}}$ will be detected to a detector D$_{j}$ is then $p(i|j)= \bra{\psi_{i}}\pi_{j}\ket{\psi_{i}}$. So if the state $\ket{\psi_{i}}$ is sent with prior probability p$_{i}$ then the probability of error is given by,
	
	\begin{equation}
		P_{err}= 1- P_{succ}  =1- \sum_{i}p_{i}\bra{\psi_{i}}\pi_{i}\ket{\psi_{i}} \label{eq:err}
	\end{equation}

	In minimum error state discrimination one aims to minimize the error probability given by Eq. \eqref{eq:err}. Although measurement stategy is available for discriminating any two nonorthogonal states for minimizing equation Eq. \eqref{eq:err} it is not in general available for three arbitrary states. Nonetheless, particular strategies are still avaiable for states possesing some relational constraints or symmetries. In ref. \cite{discrimination02}  the authers had provided a optimum strategy to discriminate three mirror symmetric quantum states. For a set of mirror symmetric states,
	
	\begin{eqnarray}
		\ket{\psi_{1}} &=& cos \theta\ket{0} + sin \theta\ket{1}, \nonumber\\  \ket{\psi_{2}} &=& cos \theta\ket{0} - sin \theta\ket{1}, \nonumber\\		\ket{\psi_{3}} &=& \ket{0}
		\label{eq:states}
	\end{eqnarray}

	\begin{figure}%
		\centering
		\subfloat[\centering Three arbitrary mirror symmetric states with $\alpha=2\theta$]{{\includegraphics[width=6cm]{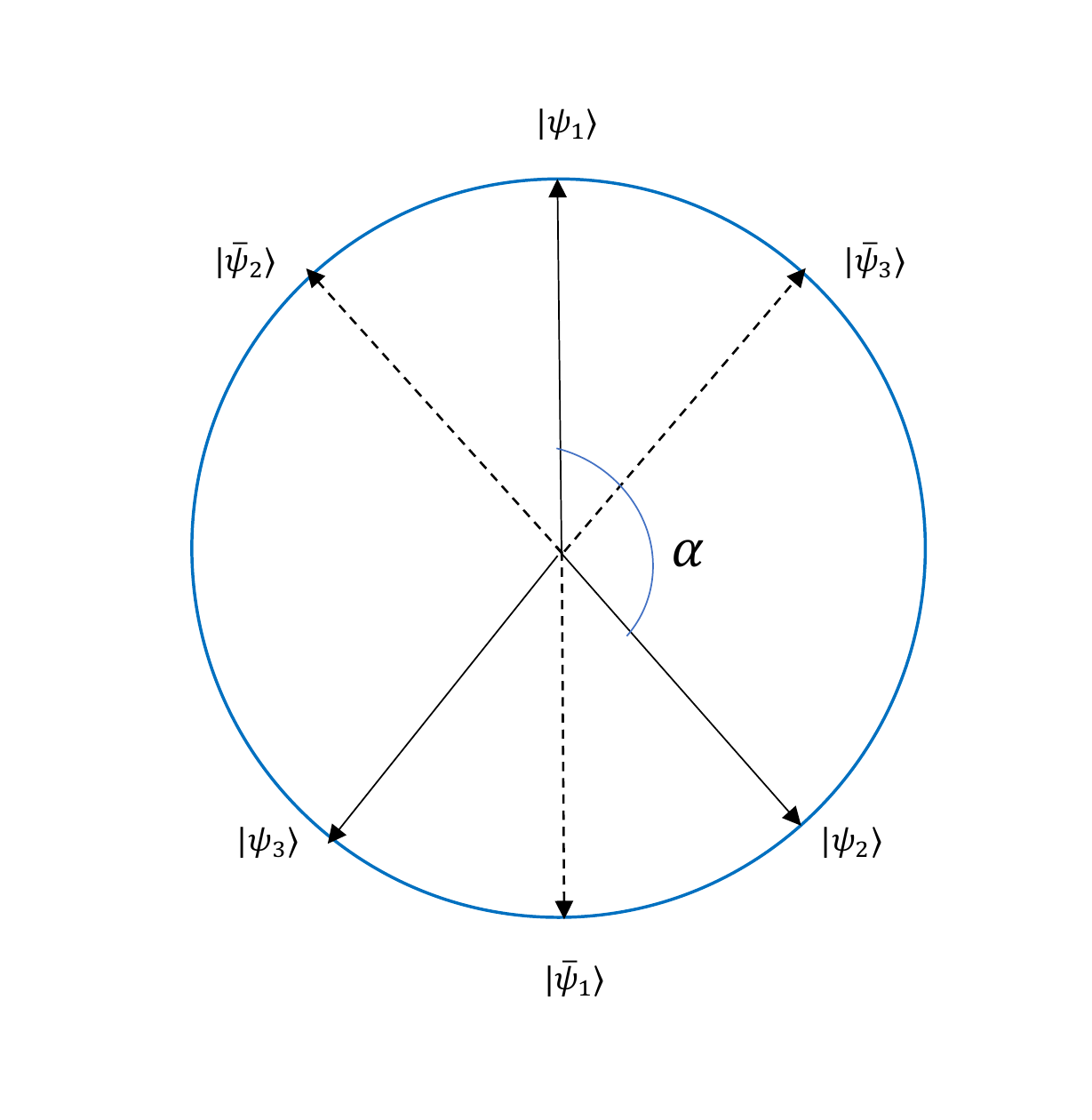} }}%
		\qquad
		\subfloat[\centering Trine state with $\theta=\pi/3$]{{\includegraphics[width=5.5cm]{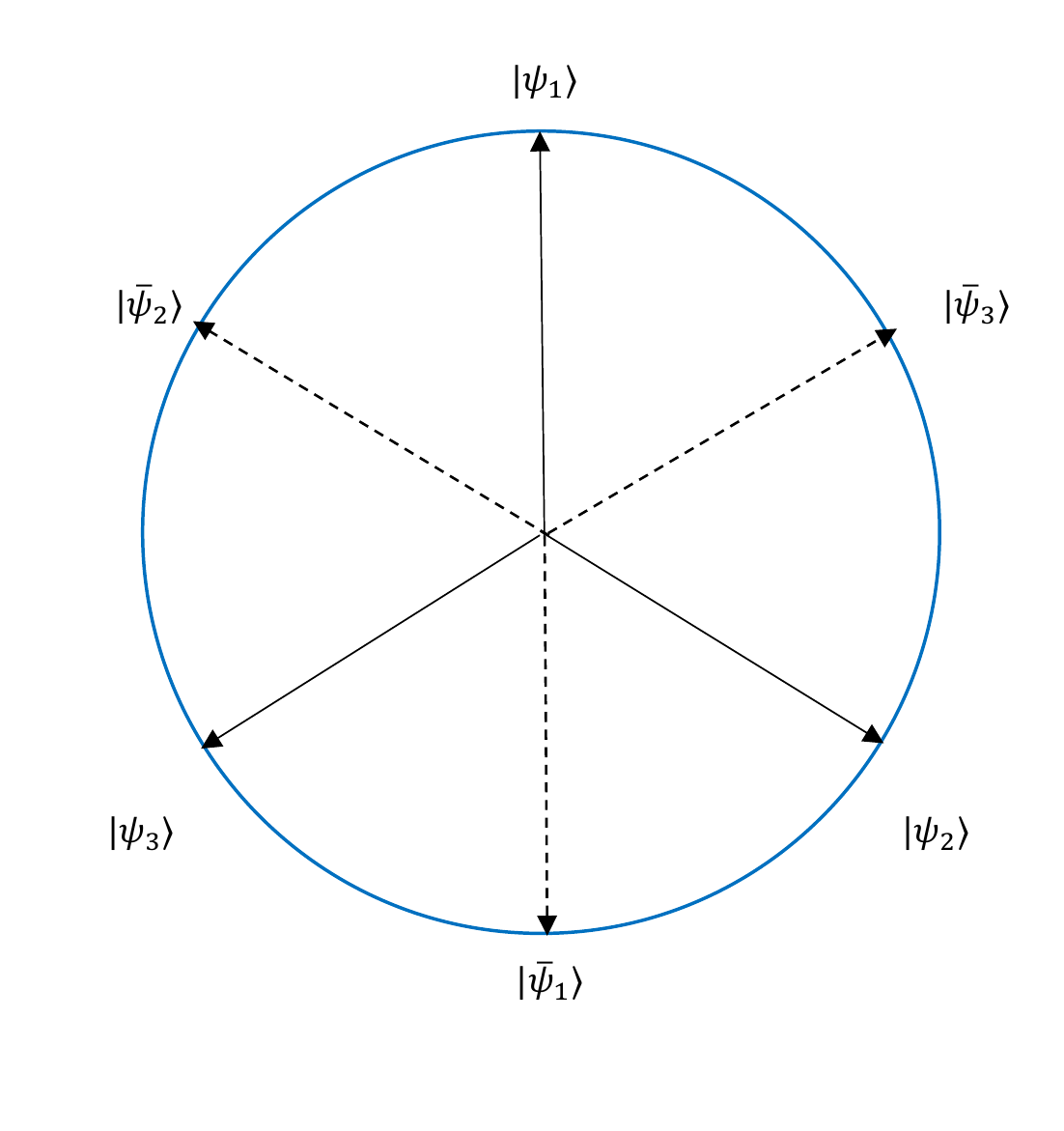} }}%
		\caption{Representation of Mirror symmetric states on the equitorial plane of the Bloch sphere}%
		\label{fig:trine}%
	\end{figure}
	
	with prior probabilities $p_{1}$= $p_{2}=p$  and $p_{3}= 1-2p$, it has been shown that employing the optimum strategy one can get quantum success probability $S_{Q}$ in accordance to the following. 	For $	p\geq 1/(2+cos\theta (cos\theta+sin\theta))$, it is

	\begin{equation}
		\begin{split}
			S_{3}=p(1+sin2\theta),
		\end{split}
	\end{equation}
	while for $p\leq 1/(2+cos\theta(cos\theta+sin\theta))$, it takes the value,
	
	\begin{equation}
		\begin{split}
			S_{3}=\frac{(1-2p)(p\sin^{2}\theta+1-2p-p\cos^{2}\theta)}{(1-2p-p\cos^{2}\theta)},
		\end{split}
	\end{equation}
	
	which reproduces the Helstrom bound \cite{helstrom} of discriminating two nonorthogonal quantum states when p$_{1}$=p$_{2}$=1/2. While as a special case with $\theta$=$\pi/3$ it gives the optimal success probabilities of discriminating between trine states.

	To find the optimum success probability of a task involving Minimum error discrimination of mirror symmetric states in a noncontextual model let us first formulate the task the ontological model. In an ontological model all the strategies are governed by the ontic state variable $\lambda$. Given three arbitrary quantum preparations $\psi_{1}$, $\psi_{2}$, $\psi_{3}$ if one tries to guess where a particular $\lambda$ comes form one can do no better than guessing one between P$(\psi_{1}|\lambda)$, P$(\psi_{2}|\lambda)$ and P$(\psi_{3}|\lambda)$ which is maximum. So in a particular run of experiment the probability that one guesses rightly is simply $max\{P(\psi_{1}|\lambda),P(\psi_{2}|\lambda),P(\psi_{3}|\lambda)\}$. Then the average optimum success probability S$_{ont}$ in an ontological model would be,

	\begin{eqnarray}
		\nonumber
		S_{3}&=&\int_{\Lambda}p(\lambda) max\{\mu(\psi_{1}|\lambda),\mu(\psi_{2}|\lambda),\mu(\psi_{3}|\lambda)\}d\lambda \label{eq:sn1}\\
		&=& \int_{\Lambda} p(\lambda)\Bigg(1- min\{\mu(\psi_{1}|\lambda),\mu(\psi_{2}|\lambda)\}\\
		\nonumber
		&-&min\{\mu(\psi_{1}|\lambda),\mu(\psi_{3}|\lambda)\}-min\{\mu(\psi_{2}|\lambda),\mu(\psi_{3}|\lambda)\}\\
		\nonumber
		&	+&min\{\mu(\psi_{1}|\lambda),\mu(\psi_{2}|\lambda),\mu(\psi_{2}|\lambda)\}\Bigg)d\lambda, 
	\end{eqnarray}
	
	where in the last step we have used the fact that if $a+b+c=1$ then $max\{a,b,c\}=1-min\{a,b\}-min\{a,c\}-min\{b,c\}+min\{a,b,c\}$. Using bayasian inversion, viz., $p(\lambda)p(\psi_{i}|\lambda)=p(\psi_{i})p(\lambda|\psi_{i})$, the Eq. \eqref{eq:sn1} can be re-written as 
	\begin{eqnarray}
		S_{3}	&=& 1-\int_{\Lambda}\Bigg( min\{\mu(\lambda|\psi_{1})p(\psi_{1}),\mu(\lambda|\psi_{2})p(\psi_{2})\}\\
		\nonumber
		&&\hspace{0cm}+min\{\mu(\lambda|\psi_{1})p(\psi_{1}),\mu(\lambda|\psi_{3})p(\psi_{3})\}\\
		\nonumber
		&&\hspace{0cm}+min\{\mu(\lambda|\psi_{2})p(\psi_{2}),\mu(\lambda|\psi_{3})p(\psi_{3})\}\\
		\nonumber
		&&\hspace{0cm}-min\{\mu(\lambda|\psi_{1})p(\psi_{1}),\mu(\lambda|\psi_{2})p(\psi_{2}),\mu(\lambda|\psi_{3})p(\psi_{3})\}\Bigg)d\lambda. \label{eq:sn2}
	\end{eqnarray}

	\begin{figure}
		\centering
		\includegraphics[scale=0.6]{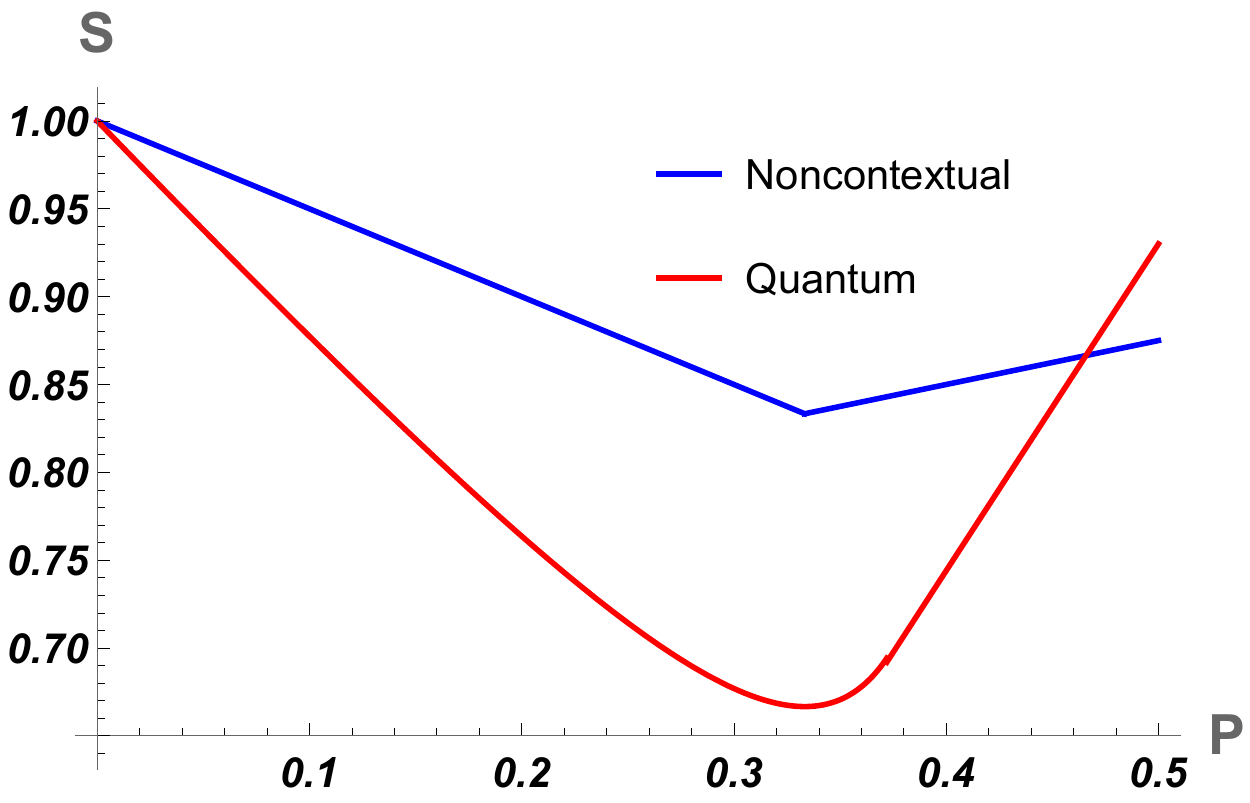}
		\caption{Comparison between nocontextual model and quantum theory in terms of success probabilities of discrimination of trine states}\label{fig1}
	\end{figure}
	
	Here $p(\psi_{i})$ denotes the prior probability of preparing state $\ket{\psi_{i}}$ with $i=1,2,3$.

	Now it is evident that,
	\begin{equation}
		\begin{split}
			\forall  |\psi_{i}\rangle, \lambda, \hspace{1cm}	min\{\mu(\lambda|\psi_{1})p(\psi_{1}),\mu(\lambda|\psi_{2})p(\psi_{2}),\mu(\lambda|\psi_{3})p(\psi_{3})\}\\\leqslant min\{\mu(\lambda|\psi_{2})p(\psi_{2}),\mu(\lambda|\psi_{3})p(\psi_{3})\}. \label{eq:less}
		\end{split}
	\end{equation}
	Impinging the inequality given by Eq. \eqref{eq:less} into Eq. \eqref{eq:sn2} we get,
	\begin{equation}
		\begin{split}
			S_{3}\leqslant 1&-\int_{\Lambda} min\{\mu(\lambda|\psi_{1})p(\psi_{1}),\mu(\lambda|\psi_{2})p(\psi_{2})\}d\lambda\\
			&-\int_{\Lambda}min\{\mu(\lambda|\psi_{1})p(\psi_{1}),\mu(\lambda|\psi_{3})p(\psi_{3})\}d\lambda\\
			\leqslant 1&-min\{p(\psi_{1}),p(\psi_{2})\}\int_{\Lambda} min\{\mu(\lambda|\psi_{1}),\mu(\lambda|\psi_{2})\}d\lambda\\
			&-min\{p(\psi_{1}),p(\psi_{3})\}\int_{\Lambda}min\{\mu(\lambda|\psi_{1}),\mu(\lambda|\psi_{3})\}d\lambda.
		\end{split}
	\end{equation}

	\begin{figure}%
		\centering
		\subfloat[\centering Top view]{{\includegraphics[width=6cm]{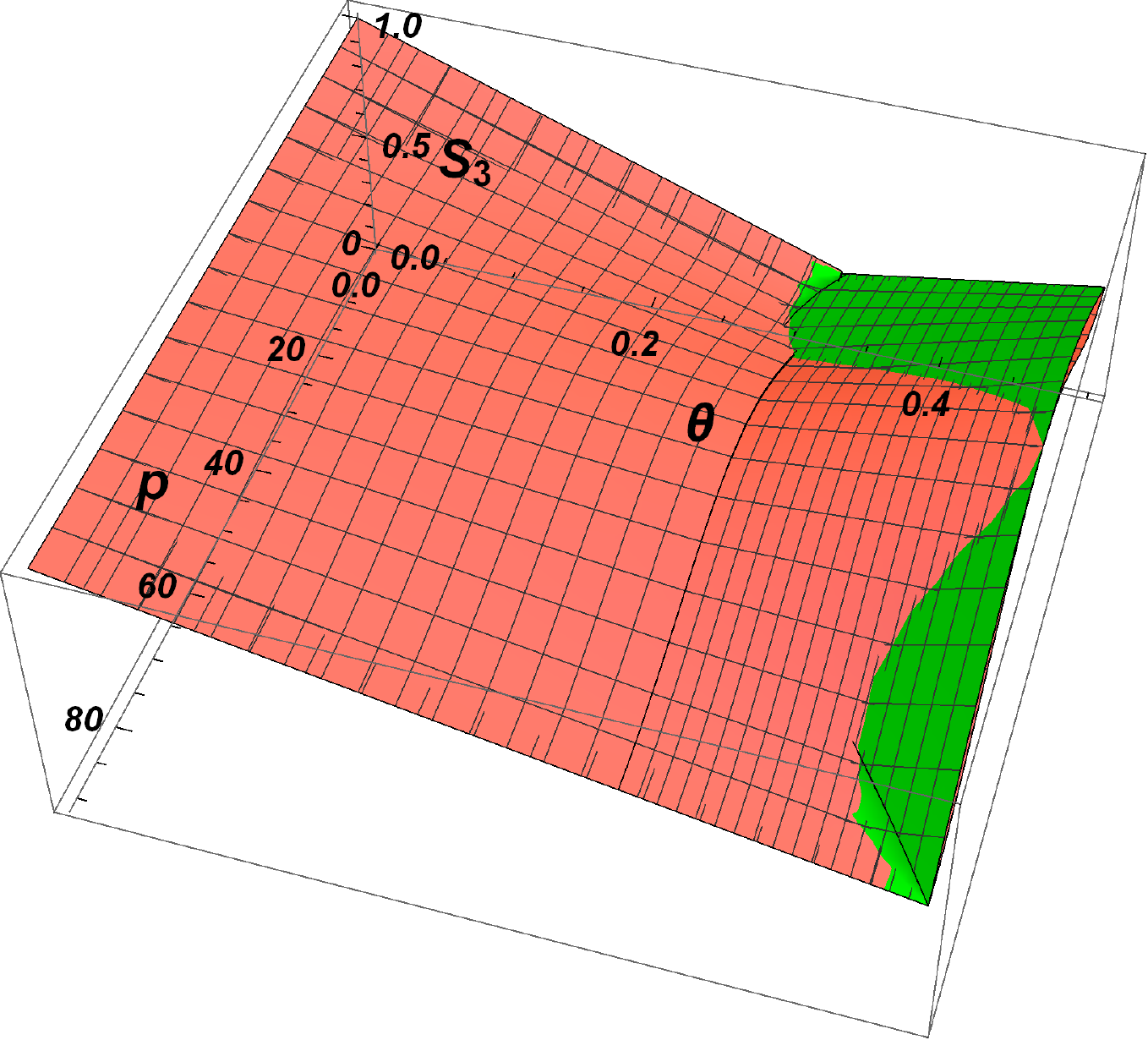} }}%
		\qquad
		\subfloat[\centering Front view]{{\includegraphics[width=6cm]{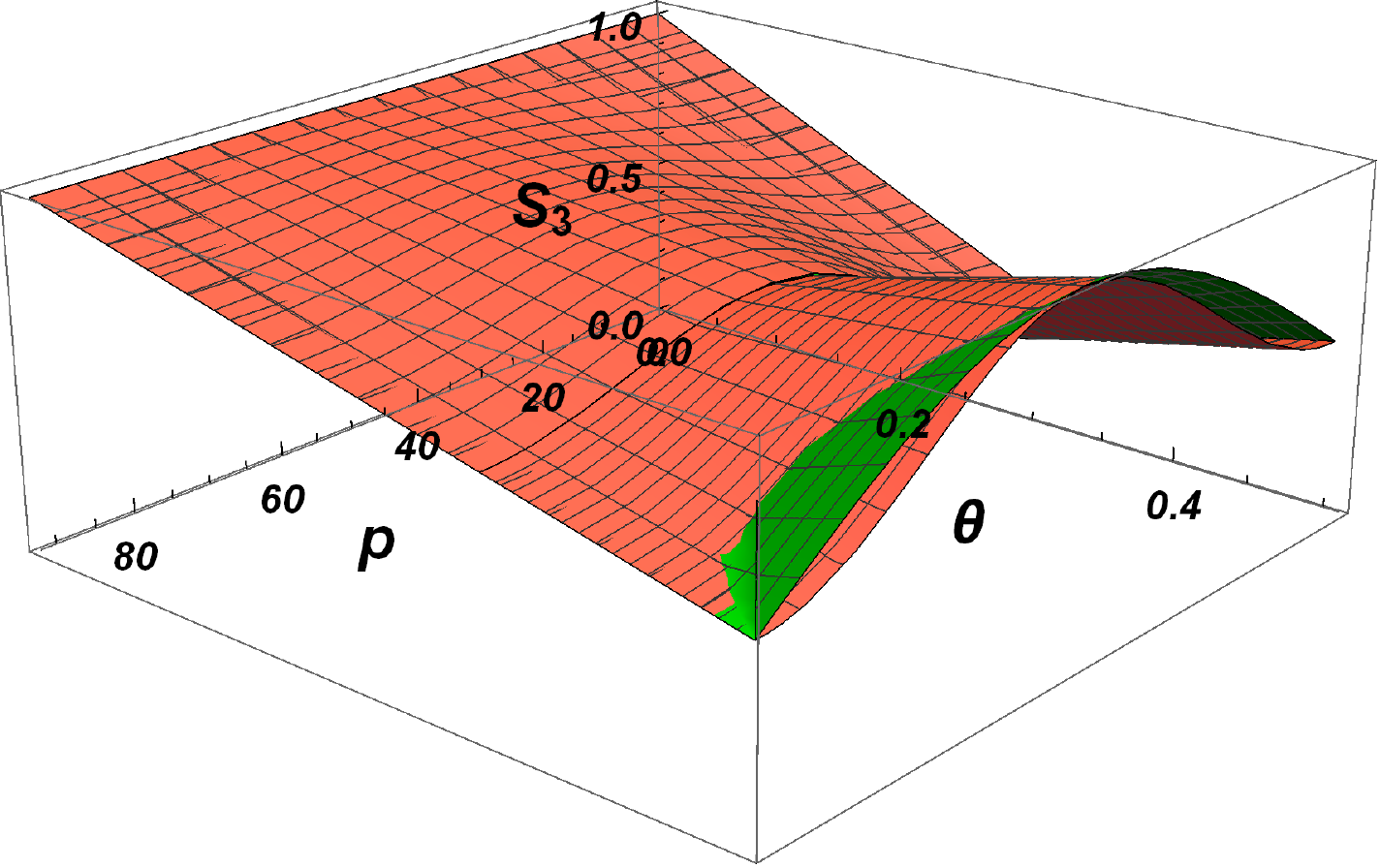} }}%
		\caption{Trade off between success probability S, prior probability of preparation p and angle $\theta$ of the supplied states in a noncontextual model (Pink) and quantum theory (Green).}%
		\label{fig:tradeoff}%
	\end{figure}
	
	As mentioned earlier, we take $p(\psi_{1})=p(\psi_{2})=p$ and $p(\psi_{3})=1-2p$. Now, for $p\leq 1/3$ we have  $min\{p,1-2p\} = p$	and for   $	p\hspace{0.1cm}\textgreater\hspace{0.1cm} 1/3$, we have $min\{p,1-2p\} = 1-2p$. Then, for $p\leq 1/3$
	\begin{equation}
		\begin{split}
			S_{3}\leqslant 1&- p\int_{\Lambda} min\{\mu(\lambda|\psi_{1}),\mu(\lambda|\psi_{2})\}d\lambda\\
			&-p\int_{\Lambda}min\{\mu(\lambda|\psi_{1}),\mu(\lambda|\psi_{3})\}. \label{eq:ont1}
		\end{split}
	\end{equation}
	Similarly for $p> 1/3$
	\begin{equation}
		\begin{split}
			S_{3}\leqslant 1&- p\int_{\Lambda} min\{\mu(\lambda|\psi_{1}),\mu(\lambda|\psi_{2})\}d\lambda\\
			&-(1-2p)\int_{\Lambda}min\{\mu(\lambda|\psi_{1}),\mu(\lambda|\psi_{3})\}d\lambda. \label{eq:ont2}
		\end{split}
	\end{equation}

	We recall that $		\int_{\Lambda}\mu(\lambda|\psi)  \xi(\psi|\lambda) d\lambda = 1$ and denote the space of $\lambda$ for which $\mu(\lambda|\psi)>0$ by $supp[\mu(\lambda|\psi)]=\Lambda_{\psi}$. Using the fact that $\int_{\Lambda}\mu(\lambda|P) d\lambda =1$, it is easy to see that, $\forall \lambda \in supp[\mu(\lambda|\psi)],\hspace{0.2cm} \xi(\psi|\lambda)=1$.

	Thus from Eq. \eqref{eq:nonc} we can write $\int_{\Lambda}min\{\mu(\lambda|\psi_{1}),\mu (\lambda|\psi_{2})\}=c^{\psi_{1},\psi_{2}}$ and $\int_{\Lambda}min\{\mu(\lambda|\psi_{1}),\mu (\lambda|\psi_{3})\}=c^{\psi_{1},\psi_{3}}$ which lead us to re-write Eq. \eqref{eq:ont1} and Eq. \eqref{eq:ont2} as,\\

	for $0 \leqslant p\leqslant 1/3$
	\begin{equation}
		\begin{split}
			S_{3}\leqslant 1-p c^{\psi_{1},\psi_{1}}-p c^{\psi_{1},\psi_{3}}
		\end{split}
	\end{equation}

	and for $1/3 < p \leqslant 1/2$
	\begin{equation}
		\begin{split}
			S_{3}\leqslant 1-pc^{\psi_{1},\psi_{2}}-(1-2p)c^{\psi_{1},\psi_{3}}.\\
		\end{split}
	\end{equation}
	Using Eq.\eqref{eq:states} we write the confusibilities as,
	\begin{equation}
		\begin{split}
			c^{\psi_{1},\psi_{2}}&=\cos^{2}2\theta,\\
			c^{\psi_{1},\psi_{3}}&=\cos^{2}\theta \label{eq:overlap}.
		\end{split}
	\end{equation}
	Finally using Eq.\eqref{eq:overlap} we get the success probability of discriminating three mirror symmetric states as,\\

	for  $0 \leqslant p\leqslant 1/3$
	\begin{equation}
		\begin{split}
			S_{3}\leqslant 1-p\cos^{2}2\theta-p\cos^{2}\theta
		\end{split} \label{eq:final1}
	\end{equation}

	and for $1/3 < p \leqslant 1/2$
	\begin{equation}
		\begin{split}
			S_{3}\leqslant 1-p\cos^{2}2\theta-(1-2p)\cos^{2}\theta
		\end{split} \label{eq:final2}
	\end{equation}

	Fig.(\ref{fig:tradeoff}) shows the trade-off between the success probability of discrimination $S_{3}$, the prior probability of state preparation p, and the characterizing angle $\theta$ of the mirror-symmetric states. The pink portion of the graph is representative of the upper bound of the success probabilities predicted by a noncontextual model whence the green portion of the graph shows the upper bound predicted by the quantum theory. It is also evident from Fig.(\ref{fig:tradeoff}) that within a restricted range of prior probabilities quantum theory provides the contextual advantage while discriminating three mirror symmetric states. This is the main result of the paper and  is summarized in the following theorem:
	
\textbf{\textit{Theorem:}} For MESD of three mirror-symmetric states, the availability of contextual advantage for a certain value of confusability is restricted by the prior probabilities with which the states are being supplied.\\

As a special case we note that for $p=1/2$,  Eq. \eqref{eq:final2} reduces to the Eq. \eqref{eq:trade1} which is the optimal trade-off between the success probability and confusibility for the discrimination of two states.

	\section{Conclusive remarks}
	
	Here we studied the preparation noncontextuality in the context of MESD. We quantified the quantum contextual advantage in terms of the success probabilities achievable in a MESD task. In this work, we first emphasized that while getting the quantum advantage in a MESD task involving more than two states, an important role is played by the prior probabilities with which the states are supplied. Prior to our work in Ref. \cite{ContxDis18} the authors have established a no-go theorem for the discrimination of two arbitrary pure quantum states. The result of \cite{ContxDis18} suggests that the quantum contextual strategy for MESD must always outperform its classical counterpart when the states are supplied with equal prior probabilities.

We extended their treatment and have shown that quantum contextual advantage is always available for any value of prior probabilities when only two states are involved in the task. Although the value of success probability for a particular value of confusability varies depending on the prior probabilities, it never goes below the success probability predicted by the noncontextual model. We then generalized it to the case when the MESD task involves three mirror-symmetric states supplied with arbitrary prior probabilities. Our result asserts that quantum theory does not always provide a contextual advantage for a particular choice of mirror-symmetric states. We have shown an interplay between the choice of states and its prior probabilities that provides a contextual advantage. We thus established the importance of the role played by the prior probabilities while discriminating more than two quantum states. For a further generalization of our work one may consider unambiguous discrimination of quantum states and study the limitations on the availability of contextual quantum advantage. Our result thus opens up an avenue to study the limitations of quantum strategies in terms of different information processing tasks.


\begin{thebibliography}{99}
		\bibitem{bell} J.S. Bell, On the Einstein Podolsky Rosen paradox, \href{https://doi.org/10.1103/PhysicsPhysiqueFizika.1.195}{Physics, {\bf 1}, 195 (1964)}.
		
		\bibitem{chsh} J. F. Clauser, M. A. Horne, A. Shimony, and R. A. Holt, Proposed Experiment to Test Local Hidden-Variable Theories, \href{https://journals.aps.org/prl/abstract/10.1103/PhysRevLett.23.880}{Phys. Rev. Lett. 23, 880 (1969)}.
		
		\bibitem{KSpeker68} S. Kochen. P. Specker, The Problem of Hidden Variables in Quantum Mechanics, \href{https://www.jstor.org/stable/24902153?seq=1#metadata_info_tab_contents} {Journal of Mathematics and Mechanics Vol. 17, No. 1 (July, 1967), pp. 59-87}.
		
		
		\bibitem{bell66} JS Bell, On the problem of hidden variables in quantum mechanics, \href{https://journals.aps.org/rmp/abstract/10.1103/RevModPhys.38.447} {Rev. Mod. Phys. 38, 447 (1966)}.
		
		
		\bibitem{harrigen} N. Harrigan and R. W. Spekkens , Incompleteness, and the Epistemic View of Quantum States, \href{https://link.springer.com/article/10.1007} {Foundations of Physics volume 40, pages125–157 (2010)}.
		
		\bibitem{spekkens05} R. W. Spekkens, Contextuality for preparations, transformations, and unsharp measurements, \href{https://journals.aps.org/pra/abstract/10.1103/PhysRevA.71.052108} {Phys. Rev. A 71, 052108 (2005)}.
		
		\bibitem{helstrom} C. W. Helstrom , Quantum detection and estimation theory, \href{https://link.springer.com/article/10.1007/BF01007479} {Journal of Statistical Physics volume 1, pages 231–252 (1969)}.
		
		
		
		\bibitem{undis1} I. D. Ivanovic, How to differentiate between non-orthogonal states, \href{https://www.sciencedirect.com/science/article/abs/pii/0375960187902222?} {Volume 123, Issue 6, 17 August 1987, Pages 257-259}.
		
		
		
		\bibitem{undis2} D.Dieks, Overlap and distinguishability of quantum states, \href{https://www.sciencedirect.com/science/article/abs/pii/0375960188908407} {Volume 126, Issues 5–6, 11 January 1988, Pages 303-306}.
		
		
		
		\bibitem{undis3} M. Ban, Error-free optimum quantum receiver for a binary pure quantum state signal, \href{https://www.sciencedirect.com/science/article/pii/0375960196001454} {Volume 213, Issues 5–6, 29 April 1996, Pages 235-238}.
		
		
		\bibitem{errdis} J. Bae, Structure of minimum-error quantum state discrimination, \href{https://iopscience.iop.org/article/10.1088/1367-2630/15/7/073037} {New J. Phys. 15 073037 (2013)}.
		
		
		\bibitem{discrimination02} E. Andersson, S. M. Barnett, C. R. Gilson, and K. Hunter, Minimum-error discrimination between three mirror-symmetric states, \href{https://journals.aps.org/pra/abstract/10.1103/PhysRevA.65.052308}{Phys. Rev. A 65, 052308 (2002)}.
		
		
		\bibitem{ContxDis18} D. Schmid and R. W. Spekkens, Contextual Advantage for State Discrimination, \href{https://journals.aps.org/prx/abstract/10.1103/PhysRevX.8.011015}{Phys. Rev. X 8, 011015  (2018)	}.
		
		
		
		\bibitem{maroni13} O. J. E. Maroney, How statistical are quantum states?, \href{https://arxiv.org/abs/1207.6906}{ arXiv:1207.6906v2 (2013)}	
		
		
		\bibitem{cavalcanti14}  J. Barrett, E. G. Cavalcanti, R. Lal, and O. J. E. Maroney, No $\psi$-Epistemic Model Can Fully Explain the Indistinguishability of Quantum States, \href{https://journals.aps.org/prl/abstract/10.1103/PhysRevLett.112.250403}{Phys. Rev. Lett. 112, 250403 (2014)}.
		
		
		\bibitem{cyril14}  C. Branciard, How 
		$\psi$
		-Epistemic Models Fail at Explaining the Indistinguishability of Quantum States, \href{https://journals.aps.org/prl/abstract/10.1103/PhysRevLett.113.020409}{Phys. Rev. Lett. 113, 020409 (2014)}.
		
		
		
		\bibitem{ballentine} L. Ballentine, Ontological Models in Quantum Mechanics: What do they tell us?, \href{https://arxiv.org/abs/1402.5689}{https://arxiv.org/abs/1402.5689v1 (2014)}.
		
		\bibitem{spekk09} R. W. Spekkens, D. H. Buzacott, A. J. Keehn, B. Toner and G. J. Pryde, Preparation contextuality powers parity-oblivious multiplexing, \href{https://journals.aps.org/prl/abstract/10.1103/PhysRevLett.102.010401}{Phys. Rev. Lett. 102, 010401(2009).}
		
		\bibitem{ghorai18} S. Ghorai and A. K. Pan, Optimal quantum preparation contextuality in an $n$-bit parity-oblivious multiplexing task, \href{https://journals.aps.org/pra/abstract/10.1103/PhysRevA.98.032110}{Phys. Rev. A, 98, 032110 (2018).} 
		
		\bibitem{pan19} A. K. Pan, Revealing universal quantum contextuality through communication games, \href{https://www.nature.com/articles/s41598-019-53701-5}{Scientific Reports volume 9, 17631 (2019). }
		
		\bibitem{saha19} D. Saha and A. Chaturvedi, Preparation contextuality as an essential feature underlying quantum communication advantage, \href{https://journals.aps.org/pra/abstract/10.1103/PhysRevA.100.022108}{Phys. Rev. A 100, 022108 (2019). }
		
		
		
		
	\end{thebibliography}
\end{document}